\documentclass[aps,prl,twocolumn,showpacs,10pt]{revtex4-1}
\usepackage{graphicx,amsmath,amssymb,amsfonts,sidecap,color}

\makeatletter
\newcommand*{\rom}[1]{\expandafter\@slowromancap\romannumeral #1@}
\makeatother

\begin{document}
 
\title{Fractional Fermions with Non-Abelian Statistics}

\author{Jelena Klinovaja}
\affiliation{Department of Physics, University of Basel,
             Klingelbergstrasse 82, CH-4056 Basel, Switzerland}
\author{Daniel Loss}
\affiliation{Department of Physics, University of Basel,
             Klingelbergstrasse 82, CH-4056 Basel, Switzerland}

\date{\today}
%73.63.Nm	Quantum wires 
%74.45.+c	Proximity effects; Andreev reflection; SN and SNS junctions
%71.10.Fd 	Lattice fermion models (Hubbard model, etc.) 
% 71.10.Pm 	Fermions in reduced dimensions (anyons, composite fermions, Luttinger liquid, etc.) (for anyon mechanism in superconductors, see 74.20.Mn)
%73.21.Cd 	Superlattices 
%73.21.Hb 	Quantum wires
% 73.21.La 	Quantum dots 
%05.30.Pr	Fractional statistics systems (anyons, etc.)
	
\begin{abstract}
We introduce a novel class of low-dimensional topological tight-binding models that allow for bound states that are fractionally charged fermions and exhibit non-Abelian braiding statistics.
The proposed model consists of a double (single) ladder of spinless (spinful) fermions in the presence of magnetic fields. We study the system analytically in the continuum limit
as well as numerically in the tight-binding representation. We find a topological phase transition with a topological gap that closes and reopens as a function of system parameters
and chemical potential. The topological phase is of the type BDI and carries two degenerate mid-gap bound states that are localized at opposite ends of the ladders. We show
numerically that these bound states are robust against a wide class of perturbations.
\end{abstract}

\maketitle

{\it Introduction.} Topological properties of condensed matter systems have attracted considerable attention in recent years. 
In particular, Majorana fermions \cite{Wilczek}, being their own antiparticles,  are expected to occur in
 a number of systems, e.g. fractional quantum Hall systems \cite{Read_2000,Nayak}, topological insulators \cite{fu, Nagaosa_2009,Ando}, optical lattices \cite{Sato,demler_2011},  $p$-wave superconductors \cite{potter_majoranas_2011}, nanowires with strong Rashba spin orbit interaction \cite{lutchyn_majorana_wire_2010, oreg_majorana_wire_2010, alicea_majoranas_2010,mourik_signatures_2012,das_evidence_2012,deng_observation_2012}, and carbon-based systems \cite{Klinovaja_CNT, bilayer_MF_2012, MF_nanoribbon}. Another class of topological systems is given by  bound states of Jackiw-Rebbi type \cite{Jackiw_Rebbi} containing fractional charge $e/2$ \cite{FracCharge_Su,FracCharge_Kivelson,FracCharge_Bell,FracCharge_Chamon,CDW,Two_field_Klinovaja_Stano_Loss_2012}. 
 Such exotic quantum states are interesting in their own right, and due to their special robustness against many forms of perturbations they offer the possibility for applications in quantum computations, especially when they exhibit non-Abelian statistics such as Majorana fermions \cite{Alicea_2012,Halperin}.

In this Letter we introduce a surprisingly simple class of  models supporting a topological phase  with  bound states that possess not only fractional charge but also exhibit non-Abelian
statistics under braiding. These bound states behave in many ways similar to the well-studied Majorana fermions in superconducting-semiconducting nanowires \cite{Alicea_2012}, but in contrast to them they are complex fermions, and quite surprisingly, emerge in the absence of superconductivity and without BCS-like pairing.

The  two non-interacting tight-binding models we propose consist of a double ladder containing spinless  particles in a uniform magnetic field and a single ladder  containing spinful particles in the presence of both  uniform  and spatially periodic magnetic fields. 
We find  a topological phase transition in these systems when varying system parameters or the chemical potential, with a characteristic closing and re-opening of a topological gap.
Inside the topological phase we find two degenerate bound states, one localized at the right and one at the left end of the system. These bound states are fractionally charged fermions and are shown
to exhibit non-Abelian braiding statistics of the Ising type. We study the systems analytically in a continuum approach, finding explicit solutions for the bound states, and confirm these findings
by independent numerics of the underlying  tight-binding model.
We further test the stability of these states numerically against a wide class of perturbations and show that the bound states are robust against most of them, except of local charge fluctuations, against
which they are partly protected by charge neutrality.

\begin{figure}[!tb]
 \includegraphics[width=\columnwidth]{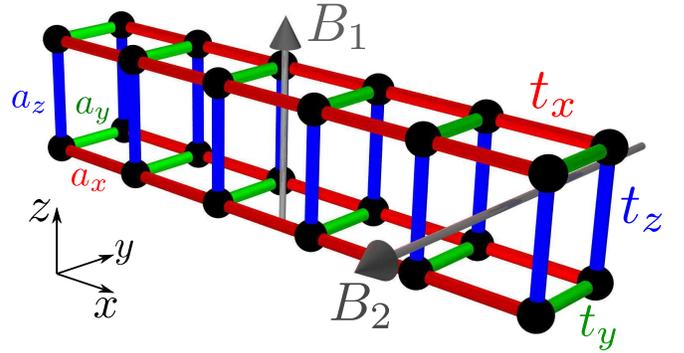}\\
% \vspace*{0.2cm} \includegraphics[width=\columnwidth,keepaspectratio=true]{tb-color.pdf}
 \caption{Double-ladder tight-binding model, consisting of a lower ($\sigma=-1$) and an upper ($\sigma=1$) ladder, each lying in the $xy$-plane and held at chemical potentials $\mu_\sigma$. 
 Here, $t_x$ (red links) is the intra-chain, $t_y$ (green links) the intra-ladder, and $t_z$ (blue links) the inter-ladder hopping amplitude, and $a_{x,y,z}$ are the corresponding lattice constants. 
 A uniform magnetic field ${\bf B} = {\bf B}_1 + {\bf B}_2$ is applied in the $yz$-plane. The associated magnetic flux results in  phases $\phi_1$ and $\phi_2$ 
 in the hopping amplitudes between 
 different chains, see Eqs. (\ref{1d_y},\ref{1d_z}).}
 \label{fig:chains}
\end{figure}

{\it Tight-binding model.} We consider a double-ladder system consisting of four coupled chains aligned along  $x$-direction,  see Fig. \ref{fig:chains}. Two upper (lower) chains form the upper (lower) ladder. Each chain is labeled by two indices $\tau$ and $\sigma$, where $\tau=\pm 1$  refers to the left/right chains, and $\sigma=\pm 1$  refers to the upper/lower ladders.  
The tight-binding Hamiltonian 
%$H_{\tau, \sigma}^x$ 
of a $(\tau, \sigma)$-chain reads
\begin{align}
&H_{\tau, \sigma}^x =  t_x\sum_{n} (c^\dagger_{\tau, \sigma, n+1}  c_{\tau, \sigma, n} +h.c.)
+\mu_{\sigma} \sum_{n} c^\dagger_{\tau, \sigma, n}  c_{\tau, \sigma, n},
\label{1d_0_tau_sigma}
\end{align}
where $c_{\tau, \sigma, n}$ ($c^\dagger_{\tau, \sigma,n} $) is the annihilation (creation) operator on  site $n$ of the $(\tau, \sigma)$-chain. The sum runs over $N$ sites composing the chain. Here, $t_x$ is the hopping matrix element in  $x$-direction, and $\mu_{\sigma}$ is the chemical potential of the $(\tau, \sigma)$-chain.

The intra-ladder coupling is given by
\begin{align}
&H^{y}_\sigma=t_y\sum_{n}( e^{i n \phi_1  } c_{1, \sigma, n}^\dagger c_{\bar{1}, \sigma, n} 
+ h.c.),
\label{1d_y}
\end{align}
where the phase 
$\phi_1$
accompanies the hopping matrix element $t_y$ in $y$-direction. This phase  arises from the magnetic flux through the unit cell produced by a magnetic field ${\bf B}_1$ in $z$-direction. 
The inter-ladder coupling between two left or two right chains  is given by 
\begin{align}
&H^{z}_{\tau}=t_z\sum_n (e^{in\phi_2}c_{\tau, 1, n}^\dagger c_{\tau, \bar{1}, n}+h.c.).
\label{1d_z}
\end{align}
Similarly, the phase  $\phi_2$ arises from the flux through the unit cell produced by a magnetic field ${\bf B}_2$  in $y$-direction. We note that in practice only one total  field, ${\bf B} = {\bf B}_1 + {\bf B}_2$, needs to be applied in the $yz$-plane, see Fig.~\ref{fig:chains}. The total tight-binding Hamiltonian for the double-ladder model is given by
\begin{equation}
H=\sum_{\tau, \sigma} [H_{\tau, \sigma}^x +\left(H^{y}_\sigma+H^{z}_{\tau}\right)/2].
\label{H_total_tb}
\end{equation}

Now we focus on a particular case of the above model. First, we fix the chemical potentials on the upper and lower ladders to be of  opposite signs, $\mu_1=-\mu_{\bar{1}}$. Second, we assume the upper ladder to be at quarter-filling, i.e. $\mu_1 =\sqrt{2} t_x$. The magnetic fields are chosen such  that  $\phi_1 =\pi/2$ and $\phi_2=\pi$, or in terms of  field strengths,
$B_1 =\Phi_0/4 a_x a_y$ and  $B_2 =\Phi_0/2 a_x a_z$,
where $\Phi_0$ is the flux quantum, and $a_{x,y,z}$ are the corresponding lattice constants. Assuming that $t_x\gg t_y, t_z$, we treat from now on inter-chain hoppings  as small perturbations. 

\begin{figure}[tb]
 \includegraphics[width=0.75\columnwidth]{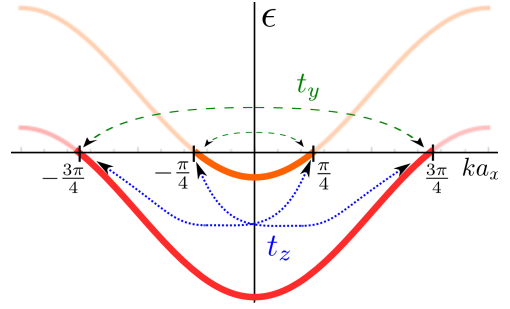}
 \caption{The spectrum of the upper  (orange) and lower   (red) chains in the first Brillouin zone. The chemical potentials are chosen  such that the system is at the charge-neutrality point. The filled (empty) states are indicated by dark-colored (light-colored) lines. The two Fermi wave vectors are given by  $k_{F,1}=\pi/4a_x$ and $k_{F,\bar1}=3\pi/4a_x$. The intra-ladder hopping $t_y$  (green dashed line) and the inter-ladder hopping $t_z$  (blue dotted lines) lead to opening of  gaps at the Fermi level ($\epsilon=0$).}
 \label{fig:spectrum}
\end{figure}

{\it Continuum model.} The most convenient way to analyze 
the tight-binding Hamiltonian in Eq. (\ref{H_total_tb}) is to go  to the continuum limit. For this we first derive the spectrum via  Fourier transformation
along the $x$-axis for each $(\tau,\sigma)$-chain,
$c_{\tau, \sigma,n}=\frac{1}{\sqrt{N}}\sum_{k}e^{i k n a_x}c_{\tau, \sigma,k}$,
where $c_{\tau, \sigma,k}$ is the annihilation operator of the electron with  momentum $k$. The Hamiltonian $H_{\tau, \sigma}^x$ has the well-known spectrum,
$\epsilon_{k,\sigma,n}=\mu_\sigma+ 2t_x\cos(ka_x)$.
At quarter-filling ($\mu_{\pm1} =\pm\sqrt{2} t_x$)  the Fermi momenta are given by $k_{F,1}=\pi/4a_x$ and $k_{F,\bar{1}}=3\pi/4a_x$, see Fig. \ref{fig:spectrum}. 
We emphasize here that the system is charge neutral, and the shifted chemical potentials just redistribute electrons between chains.

Next, we linearize the spectrum around the Fermi momenta by expressing the  annihilation operators $\Psi(x)$ that act on the states close to the Fermi level in terms of slowly varying right ($R_{\tau,\sigma}$) and left ($L_{\tau,\sigma}$)  moving fields as
\begin{equation}
\Psi(x)=\sum_{\tau,\sigma} \left[e^{ik_{F,\sigma} x}R_{\tau,\sigma}(x)+ e^{-ik_{F,\sigma} x}L_{\tau,\sigma}(x)\right].
\end{equation}
The kinetic part of the Hamiltonian corresponding to $\sum_{\tau, \sigma}H_{\tau, \sigma}^x$ [see Eq. (\ref{1d_0_tau_sigma})] is rewritten as
\begin{align}
&H^x=-i\hbar \upsilon_F\sum_{\tau,\sigma} \int dx \left(R^\dagger_{\tau,\sigma}\partial_x R_{\tau,\sigma}-L^\dagger_{\tau,\sigma}\partial_x L_{\tau,\sigma}\right),
\end{align}
where we dropped the fast oscillatory terms, and where $\upsilon_F=\sqrt{2}t_x a_x/\hbar$ is the Fermi velocity.
The intra-chain couplings, given by $H^{z}_\tau$ and  $H^{y}_\sigma$ [see Eqs. (\ref{1d_y}) and (\ref{1d_z})], lead to  mixing between $R_{\tau,\sigma}$ and $L_{\tau,\sigma}$ belonging to different chains. The intra-ladder hopping yields
\begin{align}
&H^{y}=\int dx\ t_y \left(R^\dagger_{1,1} L_{\bar{1},1}+R^\dagger_{\bar{1},\bar{1}}L_{1,\bar{1}} +h.c. \right),
\end{align}
while the  inter-ladder hopping yields
\begin{align}
&H^{z}=\int dx\ t_z \Big(R_{1, 1}^\dagger L_{1, \bar{1}}+R_{\bar{1}, 1}^\dagger L_{\bar{1}, \bar{1}}
\nonumber\\
&\hspace{65pt}
+R_{\bar{1}, \bar{1}}^\dagger L_{\bar{1}, 1}+R_{1, \bar{1}}^\dagger L_{1, 1}+h.c.\Big).
\end{align}
Next,  we introduce a new basis $\phi = (R_{1,1},L_{1,1},R_{1,\bar{1}},L_{1,\bar{1}},R_{\bar{1},1},L_{\bar{1},1},R_{\bar{1},\bar{1}},L_{\bar{1},\bar{1}})$ 
to rewrite the total Hamiltonian $H=H^x+H^y+H^z$  as 
 $H=\int dx\ \phi^\dagger(x) \mathcal{H} \phi(x)$
in terms of the Hamiltonian density $\mathcal{H}$,
\begin{align}
&\mathcal{H}=\hbar \upsilon_F \hat k \eta_3 + t_z \sigma_1\eta_1+\frac{t_y}{2} (\tau_1\eta_1-\tau_2\sigma_3\eta_2)+\delta \mu \sigma_3,
\label{eqs_ham_lin}
\end{align}
where the Pauli matrix $\eta_i$ acts on the right- and left-mover subspace, and the Pauli matrices $\sigma_i$ and $\tau_i$ act on the chain subspaces. The momentum operator  is defined as $\hat k= -i\hbar \partial_x$, with eigenvalue $k$ counted henceforth from the corresponding Fermi points.
Here, we  assume again that the chemical potentials of the upper and lower chains are opposite in sign, however, small deviations  
from  quarter-filling, $\delta \mu=\delta \mu_1=-\delta \mu_{\bar{1}}$, are taken into account. In addition, we neglect any constant shifts of the spectrum.

The Hamiltonian density $\mathcal{H}$ allows us to determine the topological class of the system \cite{Topological_class_Ludwig}. The system is invariant under the time-reversal operation $\mathcal T$, defined by  $U_T^\dagger \mathcal{H}^*(-k) U_T = \mathcal{H}(k)$. Indeed, $U_T=\tau_1\eta_1$ satisfies this relation. Similarly, the
 charge-conjugation symmetry operation $\mathcal C$, defined by $U_C^\dagger \mathcal{H}^* (-k) U_C = -\mathcal{H}(k)$, can be satisfied by $U_{C} =\tau_1\sigma_1\eta_3$. 
 Thus, 
 the system belongs to  the topological class BDI \cite{Topological_class_Ludwig}.
The one-dimensional systems of this class are allowed to have an arbitrary number of bound states inside the energy bulk gap \cite{Topological_class_Ludwig}.  To determine if bound states are present in the system for some given set of parameters, we follow the method developed in Refs. \onlinecite{MF_wavefunction_klinovaja_2012, Two_field_Klinovaja_Stano_Loss_2012}.

The eight  spectrum branches of  $\mathcal{H}$  
are given by
\begin{align}
&(\epsilon^{\pm}_{1,2})^2=(\hbar \upsilon_F k \pm \delta\mu)^2+t_z^2,\\
&(\epsilon^{\pm}_{3,4})^2=(\hbar \upsilon_F k)^2+t_z^2+t_y^2+\delta\mu^2 \nonumber \\
&\hspace{80pt}\pm 2\sqrt{t_z^2t_y^2+\delta\mu^2 [(\hbar \upsilon_F k)^2+t_y^2]}.
\end{align}
The system is gapless  only for one particular set of parameters, 
$t_y^2 =  t_z^2+\delta\mu^2$.
Otherwise, the spectrum is gapped, and there is a possibility for the existence of bound states inside this gap. Further, we are interested in states exactly in the middle of the gap, {\it i.e.} at zero energy. In addition, we focus on  semi-infinite chains, so the boundary conditions are imposed only on the left (right) end. This implies that the chain length $L$ is much larger than the localization length $\xi$ of the bound states we find.

In order to address the existence of bound states, we first find four fundamental decaying solutions of the system of linear differential equations, following from the Schrodinger equation associated with $\mathcal{H}$ [see Eq. (\ref{eqs_ham_lin})]. Second, the dimension of the null space of the corresponding Wronskian leads us to a topological criterion \cite{Two_field_Klinovaja_Stano_Loss_2012,MF_wavefunction_klinovaja_2012}  that separates a topological phase (with bound states) from a trivial phase (without bound states). We find that bound states exist provided the following {\it topological criterion} is satisfied,
\begin{equation}
t_y^2>t_z^2+\delta\mu^2.
\label{top_criterion}
\end{equation}
Working in the operator basis $\Psi(x) =(\Psi_{1,1}, \Psi_{\bar1,1}, \Psi_{\bar1,\bar1}, \Psi_{1,\bar1})$, where $\Psi_{\tau,\sigma}(x)$ is the annihilation operator on the $({\tau,\sigma})$-chain, we find the wavefunction of the state localized at the left end of the double-ladder system explicitly,
\begin{align}
&\psi_{L}(x)=\begin{pmatrix}
         i f(x)\\f^*(x)\\-i (-1)^n f^*(x)\\-(-1)^n f(x)
        \end{pmatrix},
        \label{eq:left}
\end{align}
and the wavefunction of the state localized at the right end,
$\psi_R(x)=\psi_{L}^*(L-x)$, where $n$ labels the site, $x=na_x$ \cite{footnote_odd_number}.
Here, we have introduced the notations,
\begin{align}
&f(x)=e^{i\theta/2}(e^{ik_{F,1}x}e^{-x/\xi_1+ i\delta\mu x/\hbar \upsilon_F}-e^{-ik_{F,1}x}e^{-x/\xi_2}),\nonumber\\
&e^{i\theta}=\left(\sqrt{t_y^2-\delta\mu^2}+i\delta\mu\right)/t_y.
\end{align}

\begin{figure}[tb]
 \includegraphics[width=\columnwidth]{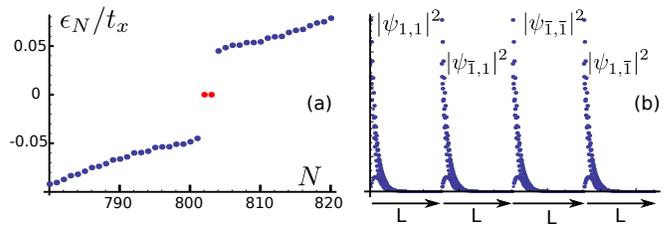}
 \caption{(a) The relevant part of the spectrum for a double-ladder system of  length $L/a_x=401$ found by numerical diagonalization of the tight-binding Hamiltonian $H$ [see Eq. (\ref{H_total_tb})]. The energy $\epsilon_N$ corresponds to the $N$th energy level.
 The parameters $t_z/t_x=0.05$, $t_y/t_x=0.1$, and $\delta \mu =0$ are chosen to satisfy the topological criterion  Eq. (\ref{top_criterion}). In the middle of the gap, $\epsilon=0$, there are two  degenerate bound states (red dots). (b) The probability density $|\psi_{\tau, \sigma}|^2$ of the left   state in the $(\tau, \sigma)$-chain. The density patterns are the same for all chains, in full agreement with the continuum solution  Eq. (\ref{eq:left}).}
 \label{fig:spectrum_state}
\end{figure}

The localization length of the bound states is given by $\xi = {\rm\ max}\{ \xi_1,\ \xi_2 \}$, where 
%the two localization lengths are given by
$\hbar \upsilon_F\xi_1=t_z$, and $\hbar \upsilon_F\xi_2=\sqrt{t_y^2-\delta\mu^2}-t_z$.
Close to the phase transition point the localization length is determined by $\xi_2$, whereas
deep inside the topological phase  by $\xi_1$. 
If $\xi$ is comparable with $L$, the two bound states localized at opposite ends overlap. As a result, the energy levels are split from zero energy, and the corresponding wavefunctions are given by the symmetric and antisymmetric combinations,
\begin{align}
\psi_{s/a} (x)=\psi_L(x_L) \pm i \psi_R(x_R).
\label{psi_+}
\end{align}
We note that these right or left localized bound states are fractionally charged fermions of charge $e/2$ \cite{FracCharge_Su, FracCharge_Kivelson, FracCharge_Bell}.

The results obtained above in the continuum model are in good agreement with the numerical results obtained by direct  diagonalization of the tight-binding Hamiltonian $H$ given in Eq. (\ref{H_total_tb}), see Fig. \ref{fig:spectrum_state}.

{\it Non-Abelian statistics.} In the absence of overlap between the bound states $\psi_L(x)$ and $\psi_R(x)$, the zero-energy level is two-fold degenerate, so it can potentially be used for braiding and ultimately for topological quantum computation. Taking into account that the system is charge-neutral, we focus on the states that have equal density distributions on both ends,
\begin{align}
&\psi_{\pm}(x) =  \psi_L(x) \pm e^{i\alpha} \psi_R(x),
% &\psi_-(x) =  \psi_L(x_L) - e^{i\alpha} \psi_R(x_R),
\label{psi_pm}
\end{align}
where $\alpha$ is an arbitrary phase (see also below). The creation operator $F_\pm^\dagger$  corresponding to the state $\psi_\pm(x)$ is given by
$F_\pm=\int dx\ \psi_\pm(x) \cdot \Psi \equiv \frac{1}{\sqrt{2}} (F_L \pm e^{i\alpha} F_R)$,
where $F_L$ ($F_R$) is the left (right) endstate annihilation operator.
During the braiding process, which corresponds to exchanging the right and left end states, the parity, defined by  $P_\pm = 1 -F_{\pm}^\dagger F_{\pm}$,  
should be conserved \cite{Halperin}.
As a result,  the old states transform into new ones as $F_{\pm} \to \pm e^{-i\alpha} F_{\pm}$ under braiding, or 
in terms of the left/right states,
\begin{align}
&F_L \rightarrow F_R,\ \ \ F_R \rightarrow e^{-2i\alpha} F_L.
\end{align}
The corresponding unitary operator $U_{LR}$ that implements this braiding rule is found to be 
$U_{LR}=e^{-i\alpha (n_++n_-)+i\pi n_-}$,
where $n_\pm = F^\dagger_\pm F_\pm$. Indeed, it is easy to show that $U_{LR}^\dagger F_\pm U_{LR}= \pm e^{-i\alpha} F_\pm$, and 
thus $U_{LR}^\dagger F_{L} U_{LR}=  F_R$, and $U_{LR}^\dagger F_{R} U_{LR}=   e^{-2i\alpha}F_L$.
In terms of the 
$F_{R/L}$ operators we have 
\begin{align}
&U_{LR}=e^{-i(\alpha -\pi/2)(n_R+n_L)-i\pi (e^{i\alpha}F_L^\dagger F_R+ h.c.)/2},
\label{U_LR}
\end{align}
where  $n_{L/R}=F^\dagger_{L/R} F_{L/R}$.
Next, let us assume a network of such double ladders similar to the one proposed for Majorana fermions \cite{Alicea_Nature}.
The braiding operations can be performed by exchanging bound states localized at different double ladders. Using $U_{LR}$ given explicitly by Eq. (\ref{U_LR}), one can show that two braiding operations do not commute, $[U_{ij}, U_{jk}]\neq 0$, where the labels $i,j,k$ denote three states. Thus, we see that our bound states  obey non-Abelian braiding statistics. 
In particular, it is interesting to consider the case of $\alpha =\pm \pi/2$, since these states, $\psi_{s/a}(x)$ [see Eq. (\ref{psi_+})], can be easily prepared by 
lifting the degeneracy temporarily (via tuning the wave 
function overlap), resulting in filling either the symmetric or the antisymmetric energy levels after some dephasing time. Moreover, we can determine the phase $\alpha$ [see Eq. (\ref{psi_pm})] by projective measurements on the states $\psi_{s/a}$. For example, the probability to measure  the symmetric state is given by $|\left\langle \psi_{s} | \psi_{+} \right\rangle|^2 = \sin^2 (\alpha/2 + \pi/4)$.
All this taken together opens up the possibility to use these bound states for topological quantum computation along the lines proposed for Majorana fermions  \cite{Alicea_Nature}.

\begin{figure}[!tb]
 \includegraphics[width=\columnwidth]{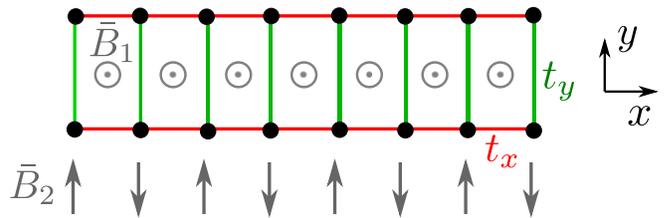}
 \caption{Single-ladder model. A uniform magnetic field $\bar{{\bf B}}_1$ is applied perpendicular to the ladder. A spatially periodic magnetic field $\bar{{\bf B}}_2$ with period $2a_x$ is applied in the $xy$-plane.}
 \label{fig:one_ladder}
\end{figure}

{\it Single-ladder model.} An alternative representation of the double-ladder model with spinless particles is a single-ladder model with spinful particles, see Fig.  \ref{fig:one_ladder}. In this model the $\sigma=-1$ (lower) and the $\sigma=1$ (upper) ladders are identified with the spin-down and spin-up components, resp.
A uniform magnetic field $\bar{{\bf B}}_1$ is applied perpendicular to the ladder, leading to  both orbital and spin effects. The  Zeeman energy acts as a chemical potential with opposite signs for opposite spin directions, $\mu_\sigma = \sigma g\mu_B \bar{B}_1 $, where $g$ is the $g$-factor, and $\mu_B$ is the Bohr magneton.
A spatially periodic field $\bar{{\bf B}}_2$ of period $2a_x$ couples opposite spins, so the effective hopping is given by
$t_z = g\mu_B \bar{B}_2$.
Such a periodic field can be produced by nanomagnets \cite{exp_field} 
 or, equivalently, by Rashba spin orbit interaction and a uniform magnetic field \cite{Braunecker_Loss_Klin_2009}.

{\it Stability against perturbations.}
We next address the stability of the topological phase and the bound states against local perturbations. In general,
we find numerically that local fluctuations that preserve the symmetry between the upper and lower ladders are not harmful. This includes  correlated fluctuations of  chemical potentials
($\Delta \mu_1=-\Delta \mu_{\bar1}$), magnetic fluxes, and hopping matrix elements. 
If the symmetry is not preserved, the perturbations act like a level-detuning, and the bound states separate in energy independent of wave function overlap but proportional to their occupation probability
at the site of fluctuation. We emphasize that the single-ladder model is protected against such symmetry-breaking terms, except against  local chemical potential fluctuations
 that break  charge neutrality, i.e. $\Delta \mu_1=\Delta \mu_{\bar1}$.
However, we note that  chains  without charge impurities are stable against such fluctuations, as
local differences in $\mu$ would lead to charge redistribution, restoring a uniform chemical potential in the chain.
Another problem can arise from flux fluctuations. They can decrease the Fourier components at $k_{F,\sigma}$ of the backscatterig terms and thereby reduce the gaps. As a result, the system can move out of the topological phase. However, these flux fluctuations become irrelevant deep inside the topological phase.

{\it Conclusions.} We have uncovered model systems of striking simplicity that allow for a topological phase with degenerate bound states that are fractionally charged and obey non-Abelian braiding statistics.
We have shown that these exotic quantum states are rather robust against a large class of perturbations.  
Quite remarkably, our models demonstrate that non-Abelian states can exist in  single-particle systems, without any correlations and in the complete absence of superconductivity or BCS-like pairing. 
This should open the path
for novel implementations of topological matter in realistic systems. One promising candidate system
that suggests itself for implementations of such tight-binding ladders
are optical lattices \cite{optical_lattice}, because they allow for a high
degree of control and, in particular, possess the charge stability of the type invoked here.

This work is supported by the Swiss NSF, NCCR Nanoscience, and NCCR QSIT.


\begin{thebibliography}{100}
\bibitem{footnote_odd_number} Here, we assume that the chains are composed of an odd number of sites. The even number case is obtained by replacing $(-1)^n$ by $(-1)^{n+1}$ in $\Psi_R(x)$.%
\end{thebibliography}
\end{document}